\documentclass[12pt]{article}
\usepackage{graphicx}

\newcommand\pubnumber{DPF2015-181\\
UCHEP-15-04}
\newcommand\pubdate{\today}

\newcommand\pubblock{\rightline{\begin{tabular}{l} \pubnumber\\
         \pubdate  \end{tabular}}}

\newenvironment{Presented}{\begin{quotation} \begin{center} 
             PRESENTED AT\end{center}\bigskip 
      \begin{center}\begin{large}}{\end{large}\end{center} \end{quotation}}

\def\beq{\begin{equation}}
\def\eeq#1{\label{#1}\end{equation}}
\def\eeqn{\end{equation}}

\def\beqa{\begin{eqnarray}}
\def\eeqa#1{\label{#1}\end{eqnarray}}
\def\eeqan{\end{eqnarray}}

\let\bar=\overbar

\def\Dslash{\not{\hbox{\kern-4pt $D$}}}
\def\dslash{\not{\hbox{\kern-2pt $\del$}}}

\def\msb{{\bar{\ssstyle M \kern -1pt S}}}

\begin{document}
\begin{titlepage}
\pubblock

\vfill
{\bf\boldmath\huge
\begin{center}
Recent results on charmless hadronic $B$ decays at Belle 
\end{center}
}
\vspace*{2.0cm}

\begin{center}
Bilas Pal\footnote{palbs@ucmail.uc.edu}, University of Cincinnati\\
On behalf of the Belle collaboration\\
\end{center}

\vspace{\fill}
\begin{abstract}
  \noindent
Two-body charmless hadronic decays of $B$ mesons are
important for determining Standard Model  parameters
and for detecting the presence of new physics. We present recent results from the Belle experiment on the charmless
hadronic decays  $B\rightarrow \eta \pi^0$ and $B\rightarrow \pi^0 \pi^0$. 
\end{abstract}
\vfill
\begin{Presented}
DPF 2015\\
The Meeting of the American Physical Society\\
Division of Particles and Fields\\
Ann Arbor, Michigan, August 4--8, 2015\\
\end{Presented}
\vfill
\end{titlepage}
\def\thefootnote{\fnsymbol{footnote}}
\setcounter{footnote}{0}

\section{Introduction}
Two-body charmless hadronic decays of $B$ mesons are
important for determining Standard Model  parameters
and for detecting the presence of new physics. We present recent results from the Belle experiment on the  charmless
hadronic decays  $B\rightarrow \eta \pi^0$ and $B\rightarrow \pi^0 \pi^0$. 

\section{Evidence for the decay $B\rightarrow \eta \pi^0$}
\begin{figure}[htb]
\begin{center}
\includegraphics[width=0.4\textwidth]{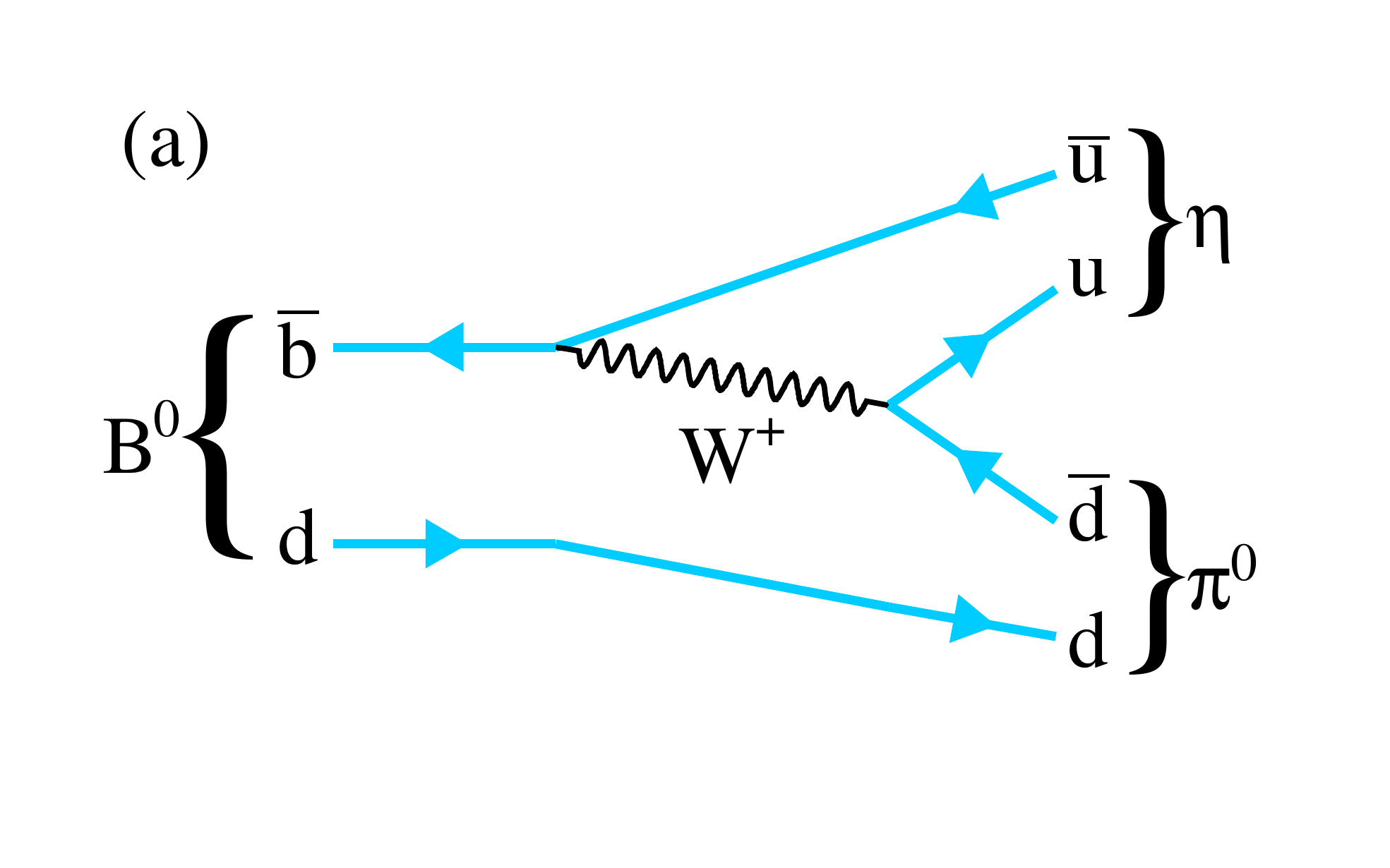}%
\includegraphics[width=0.4\textwidth]{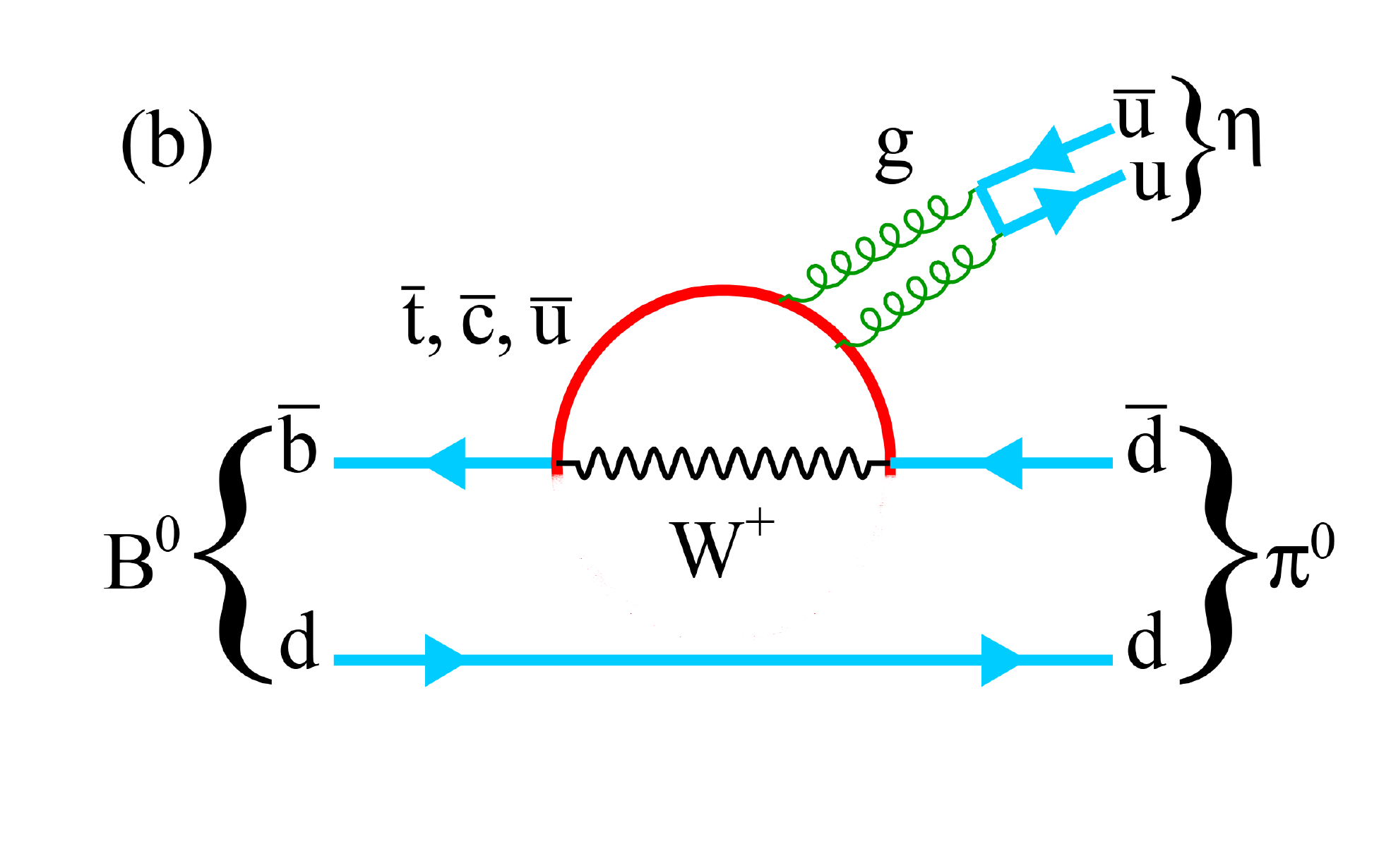}
\vskip -0.75cm
\caption{\small (a) Tree  and (b) penguin diagram 
contributions to  $B\rightarrow \eta \pi^0$ . }
\label{fig:feynman}
\end{center}
\end{figure}

The decay $B\rightarrow \eta \pi^0$  proceeds mainly via a $b\to u$ Cabibbo- and
color-suppressed ``tree'' diagram, and via a $b\to d$ 
``penguin" diagram, as shown
in Fig.~\ref{fig:feynman}. 
The branching fraction can
be used to constrain isospin-breaking effects on the value of
$\sin2\phi_2~(\sin 2\alpha)$  measured in $B\to\pi\pi$
decays~\cite{Gronau:2005pq,Gardner:2005pq}.
It can also be used
to constrain  $CP$-violating parameters ($C^{}_{\eta' K}$
and $S^{}_{\eta' K}$) governing the time dependence of
$B^0\to\eta' K^0$ decays~\cite{etapKs_bound}.
The branching fraction is estimated 
using QCD factorization~\cite{qcd}, soft collinear effective
field theory~\cite{Williamson:2006hb}, and flavor SU(3) symmetry~\cite{su3}
and is found to be in the range $(2 - 12)\times10^{-7}$.

Several experiments~\cite{Albrecht:1990am, Acciarri:1995bx, Richichi:1999kj, Chang:2004fz, Aubert:2008fu}, including Belle, have searched for this decay mode.
The current most stringent limit on the branching fraction
is $\mathcal{B}(B^0\to\eta\pi^0)<1.5\times10^{-6}$ at 
90\% confidence level (C.L.)~\cite{Aubert:2008fu}. The analysis presented here uses  the full data set of the Belle experiment running on
the $\Upsilon(4S)$ resonance at the KEKB asymmetric-energy 
$e^+e^-$ collider. This data set corresponds to
$753\times10^{6}$ $B\overline{B}$ pairs, which is a factor of 5
larger than that used previously. Improved tracking, photon reconstruction,
and continuum suppression algorithms are also used in this analysis.

\begin{figure}[h!t!p!]
\begin{center}
    \includegraphics[width=0.32\textwidth]{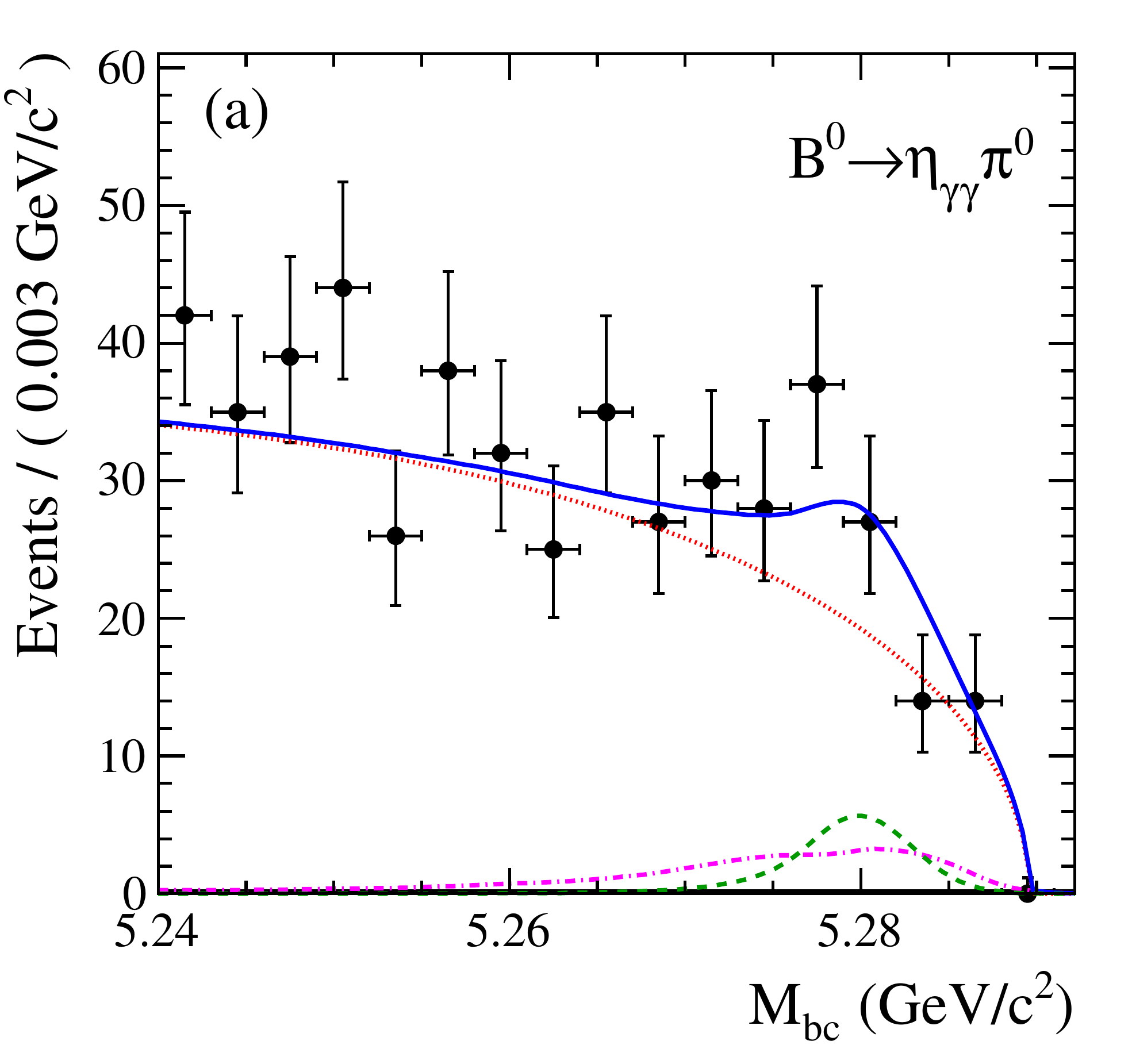}%
    \includegraphics[width=0.32\textwidth]{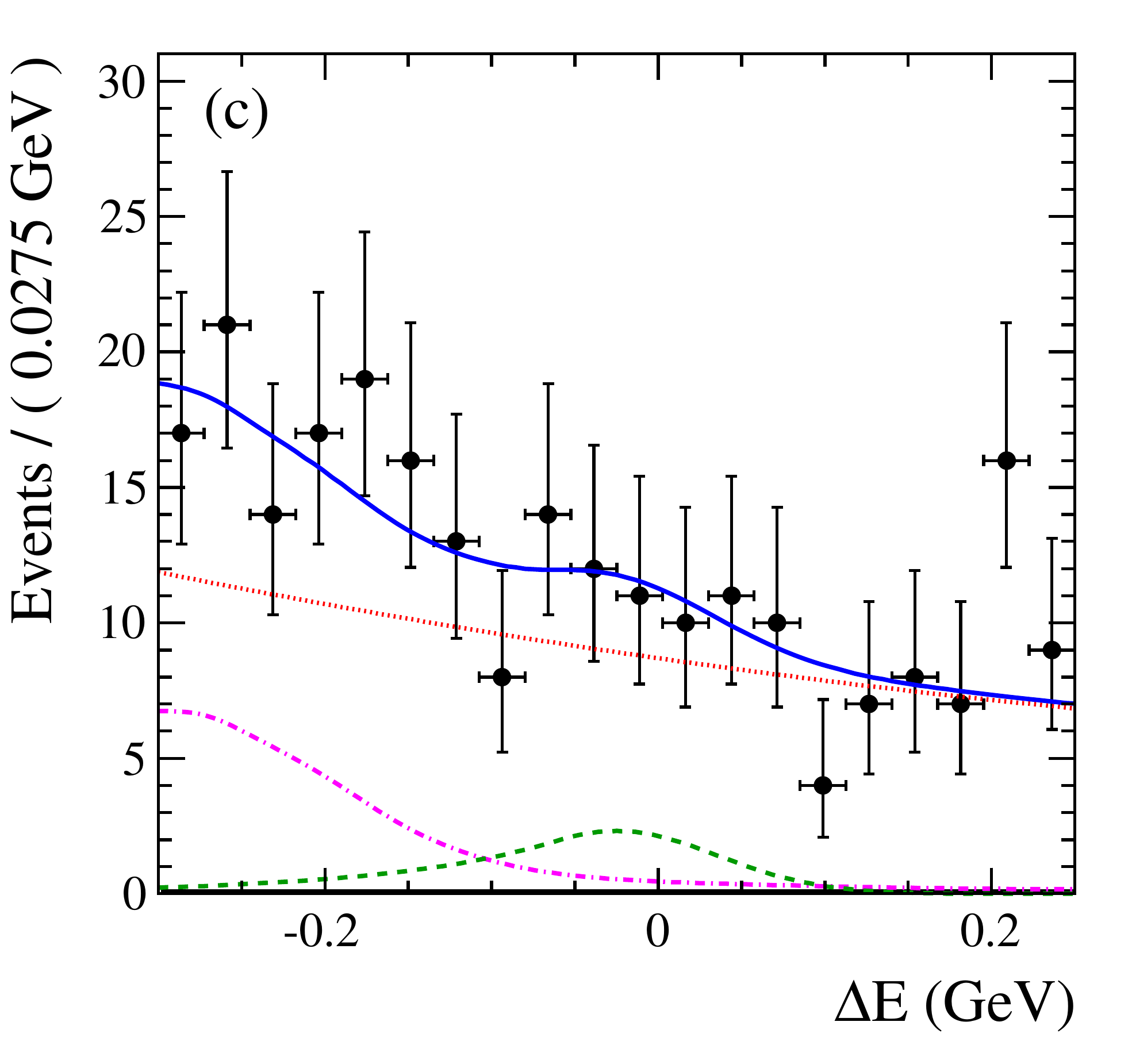}%
    \includegraphics[width=0.32\textwidth]{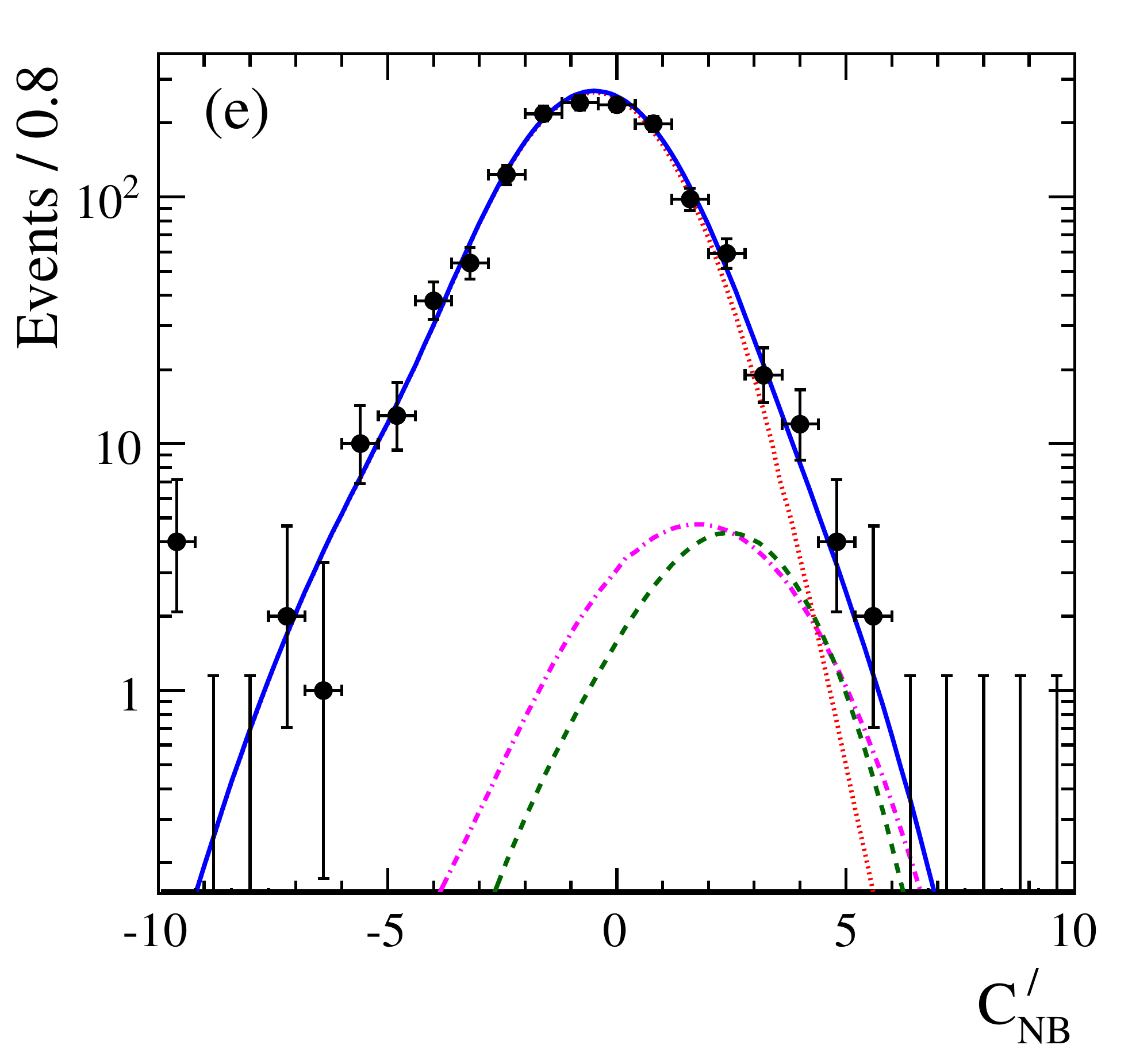}  
    \includegraphics[width=0.32\textwidth]{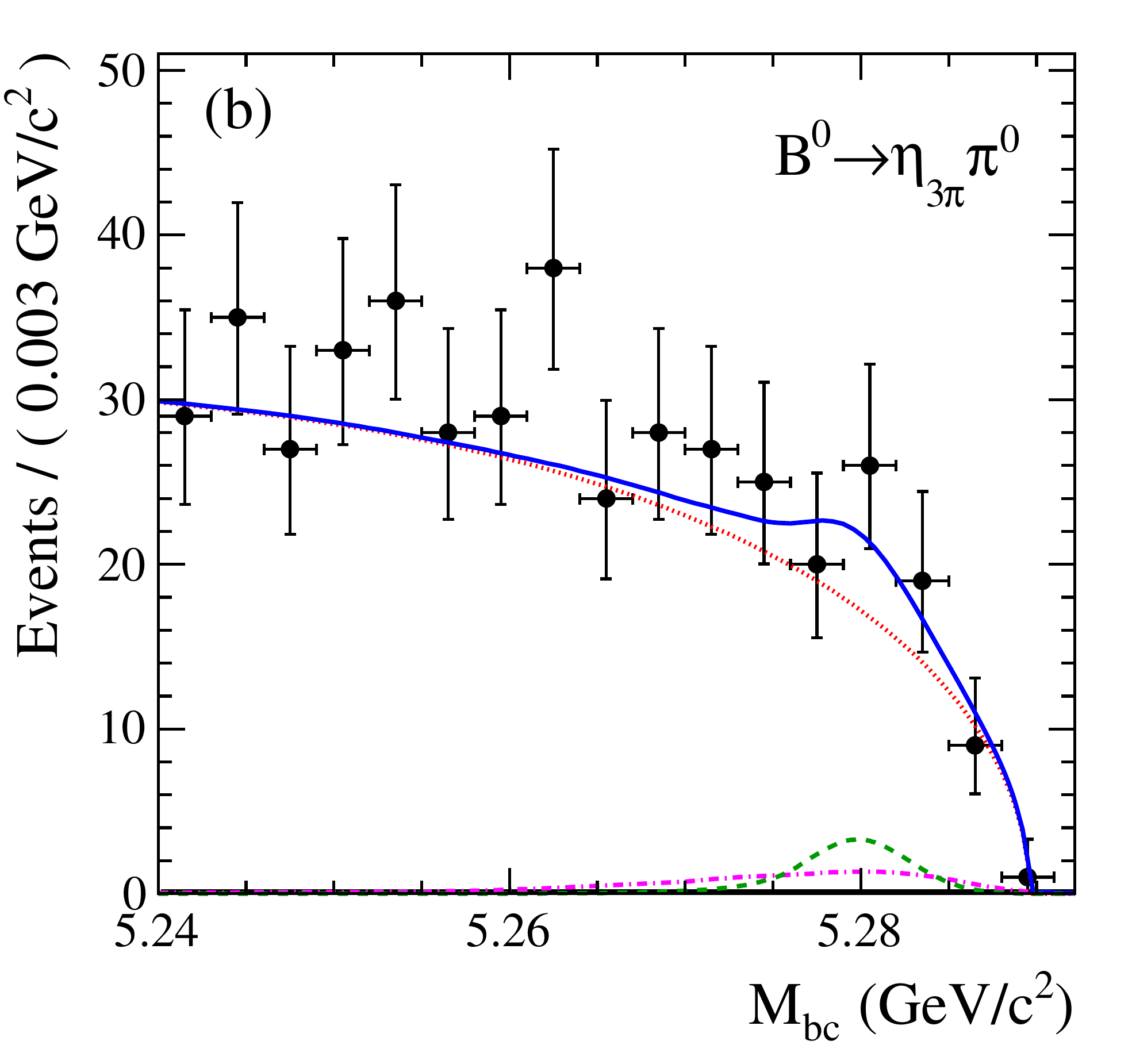}%
   \includegraphics[width=0.32\textwidth]{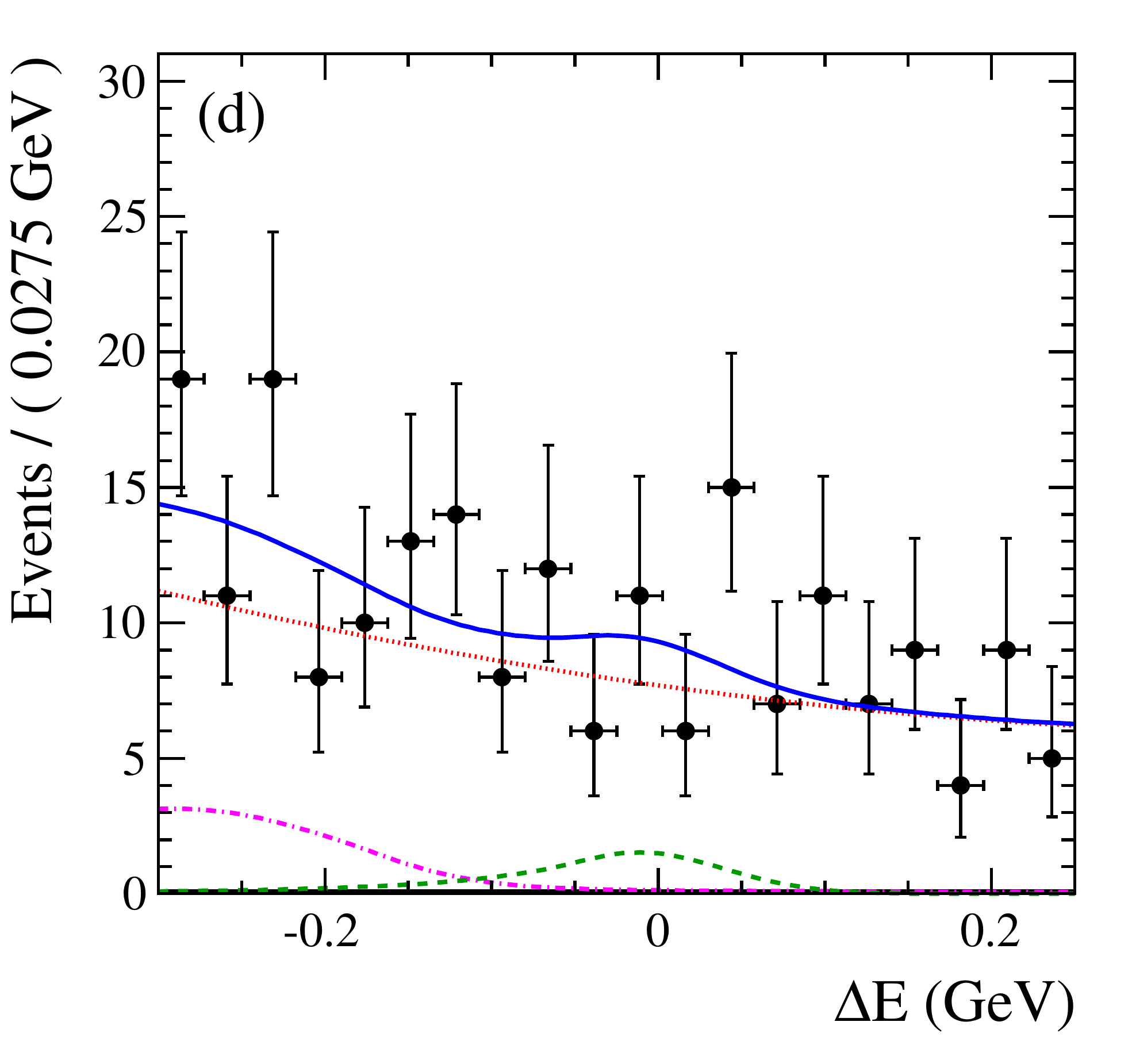}%
    \includegraphics[width=0.32\textwidth]{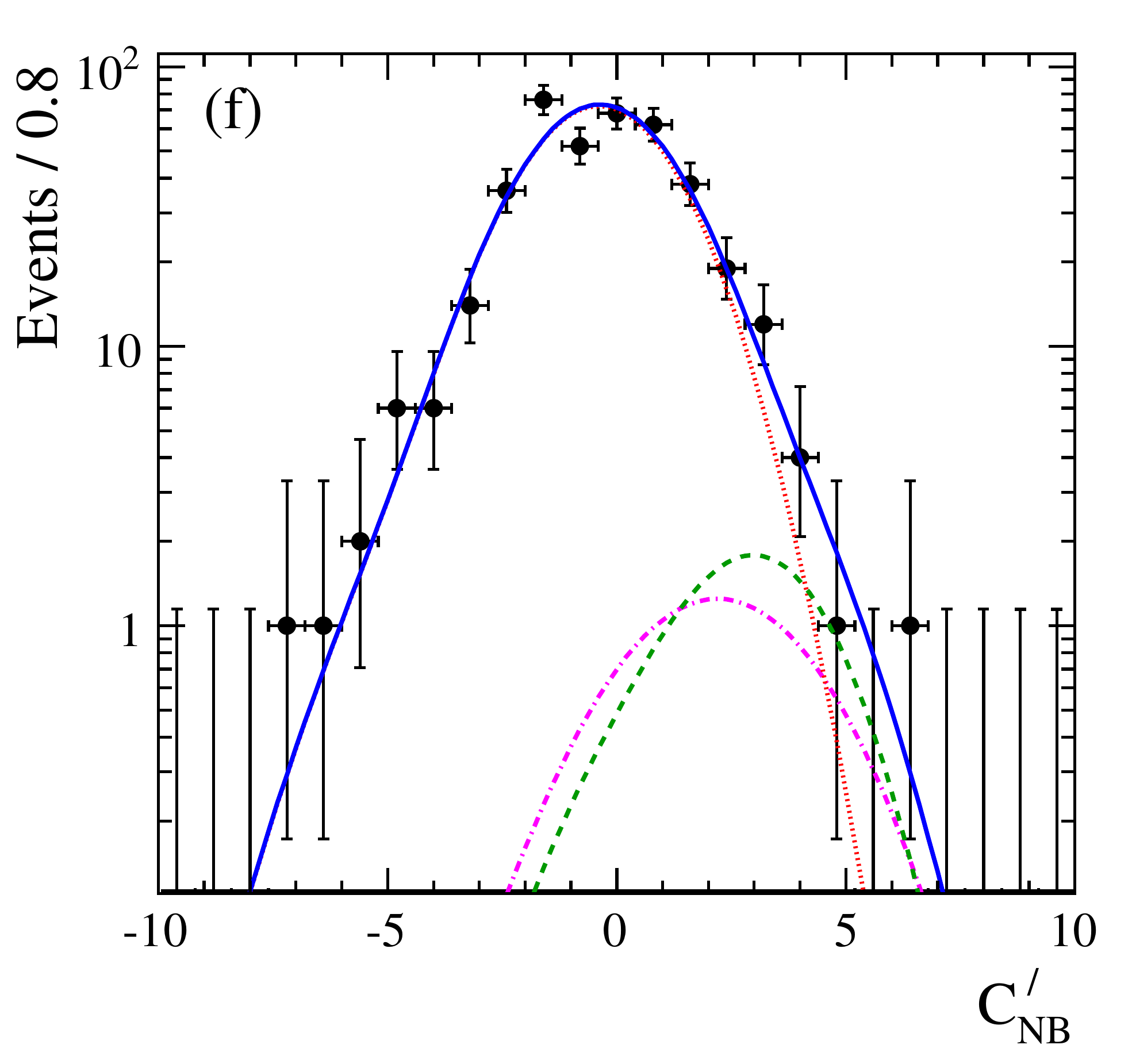}
\end{center}
\vskip -0.5cm
\caption{\small Signal enhanced  projections of the simultaneous fit for the decay $B^0\rightarrow\eta\pi^0$:
(a), (b) $M_{\rm bc}$; (c), (d) $\Delta E$;
(e), (f) $C'_{NB}$.
The top (bottom) row corresponds to $\eta\to\gamma\gamma$
($\eta\to\pi^+ \pi^-\pi^0$) decays. 
Points with error bars are data;  the (green) dashed, (red) dotted and  (magenta) dot-dashed  curves represent the signal,  continuum and charmless rare  backgrounds, respectively, and the (blue) solid curves represent
the total PDF.}
\label{fig:real_full}
\end{figure}

We find the evidence of the decay $B\rightarrow \eta \pi^0$~\cite{Pal:2015ewa}, where the candidate $\eta$ mesons
are reconstructed via $\eta\rightarrow\gamma\gamma~(\eta_{\gamma\gamma})$ and $\eta\rightarrow\pi^+\pi^-\pi^0~(\eta_{3\pi})$ decays and 
$\pi^0$ via $\pi^0\rightarrow\gamma\gamma$. Results of the fit to the variables, beam-energy-constrained mass  $M_{\rm bc}=\sqrt{E^2_{\rm beam}-|\vec{p_B}|^2c^2}/c^2$, energy difference $\Delta E=E_B-E_{\rm beam}$ and  continuum suppression variable $C'_{NB}=\ln(\frac{C_{NB}-C^{\rm min}_{NB}}{C^{\rm max}_{NB}-C_{NB}})$,  are given in Table.~\ref{fit}.
\begin{table}[htb]
\begin{center}
\renewcommand{\arraystretch}{1.5}
\caption{\small Fitted signal yield $Y_{\rm sig}$,
reconstruction efficiency $\epsilon$,
$\eta$ decay branching fraction $\mathcal{B}_{\eta}$,
signal significance, and $B^0$ branching fraction 
$\mathcal{B}$ for the decay $B^0\rightarrow\eta\pi^0$. The errors listed are statistical only.
The significance  includes both statistical and
systematic uncertainties.}
\label{fit}
\begin{tabular}{c|ccccc}
\hline \hline
Mode & $Y_{\rm sig}$ & $\epsilon(\%)$ & $\mathcal{B}_{\eta}(\%)$ & 
Significance & $\mathcal{B}(10^{-7})$\\
\hline
$B^0\to\eta_{\gamma\gamma}\pi^0$ & $30.6^{+12.2}_{-10.8}$ &
       18.4 & 39.41 & 3.1 & $5.6^{+2.2}_{-2.0}$ \\
$B^0\to\eta_{3\pi}\pi^0$ & $0.5^{+6.6}_{-5.4}$ &
       14.2 & 22.92 & 0.1 & $0.2^{+2.8}_{-2.3}$ \\
Combined & & & & 3.0 & $4.1^{+1.7}_{-1.5}$ \\
\hline\hline
\end{tabular}
\end{center}
\end{table}
The combined branching fraction
is determined by simultaneously fitting both $B^0\to\eta_{\gamma\gamma}\pi^0$
and $B^0\to\eta_{3\pi}\pi^0$ samples for a common  $\mathcal{B}(B^0\to\eta\pi^0)$.  Signal enhanced projections of the simultaneous fit are
shown in Fig.~\ref{fig:real_full}.
The branching
fraction for $B\rightarrow \eta \pi^0$  decays is measured to be
\begin{eqnarray}
\mathcal{B}(B^0\to\eta\pi^0) & = & 
\left( 4.1^{+1.7+0.5}_{-1.5-0.7}\right) \times 10^{-7}\nonumber,
\end{eqnarray}
where the first uncertainty is statistical and the second
is systematic. This corresponds to a 90\% C.L. upper limit of $\mathcal{B}(B^0\to\eta\pi^0)<6.5\times 10^{-7}$. The significance of this result is~$3.0$ standard deviations.
The measured branching fraction  is in good agreement with theoretical
expectations~\cite{qcd, Williamson:2006hb, su3}. Inserting our measured value
into Eq. (19) of Ref.~\cite{Gronau:2005pq} gives the result that the
isospin-breaking correction  to the  weak phase $\phi_2$ measured in $B\to\pi\pi$ decays due to $\pi^0$--$\eta$--$\eta'$ mixing is less than $0.97^{\circ}$ at 90\% C.L.
\section{The decay $B^0\rightarrow\pi^0\pi^0$ (preliminary results)}
\begin{figure}[h!t!p!]
\begin{center}
    \includegraphics[width=0.98\textwidth]{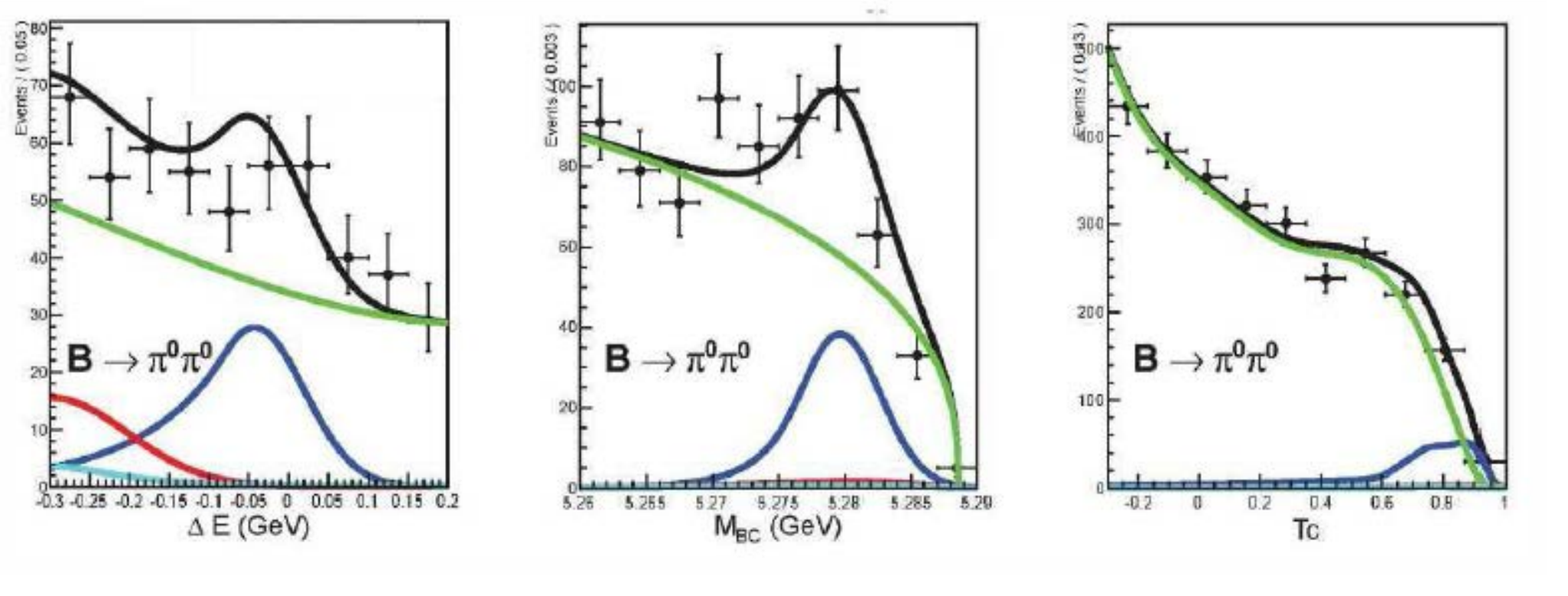}%
\end{center}
\vskip -0.5cm
\caption{\small Signal enhanced  projections of the  fit for the decay $B^0\rightarrow \pi^0\pi^0$:
(left) $\Delta E$, (middle) $M_{\rm bc}$ and (right) $T_C$.
Contributions from signal, continuum, $\rho\pi^+$ and other $B$ decays are shown by blue, green, red and cyan curves respectively.}
\label{fig:real_full2}
\end{figure}

This decay is an important input for the isospin analysis in the $B\rightarrow\pi\pi$ system. 
A fit to the variables $\Delta E$, $M_{\rm bc}$ and a fisher discriminant $T_C$ is performed. We measure  a preliminary branching fraction of 
$\mathcal{B}(B^0\to\pi^0\pi^0)=(0.9\pm0.12( {\rm stat.}) \pm0.10({\rm sys.}))\times10^{-6}$ , with a significance of 6.7 standard deviations
and the direct $CP$ asymmetry of $A_{CP}=-0.054\pm0.086$. Signal enhanced projections are shown in Fig.~\ref{fig:real_full2}. With this result, the constraint to the $\phi_2$ using the isospin relation in the $B\rightarrow\pi\pi$ system will be re-evaluated.
\section{Summary}
Using the full set of Belle data, recent and preliminary  measurements of charmless hadronic $B$ decays are presented. Our measurement of $B^0\rightarrow\eta\pi^0$ branching fraction constitutes the first evidence of the decay.
\section*{Acknowledgements}
\noindent
The author thanks the organizers of DPF 2015 for  excellent hospitality and for assembling a nice scientific program.
The author would also  like to thank  Alan Schwartz for  reviewing and providing valuable feedback to the final manuscript. This work is supported by the U.S. Department of Energy.


\begin{thebibliography}{99}


\bibitem{Gronau:2005pq} 
  M.~Gronau and J.~Zupan,
  Phys.\ Rev.\ D {\bf 71}, 074017 (2005).

\bibitem{Gardner:2005pq} 
  S.~Gardner,
  Phys.\ Rev.\ D {\bf 72}, 034015 (2005).

\bibitem{etapKs_bound}
  M.~Gronau, J.~L.~Rosner and J.~Zupan,
  Phys.\ Lett.\ B {\bf 596}, 107 (2004);
  Phys.\ Rev.\ D {\bf 74}, 093003 (2006).

\bibitem{qcd}
  M.~Z.~Yang and Y.~D.~Yang,
  Nucl.\ Phys. {\bf B609}, 469 (2001);
  M.~Beneke and M.~Neubert,
  Nucl.\ Phys. {\bf B675}, 333 (2003);
  J.~f.~Sun, G.~h.~Zhu and D.~s.~Du,
  Phys.\ Rev.\ D {\bf 68}, 054003 (2003);
  Z.~j.~Xiao and W.~j.~Zou,
  Phys.\ Rev.\ D {\bf 70}, 094008 (2004);
  H.~s.~Wang, X.~Liu, Z.~j.~Xiao, L.~b.~Guo and C.~D.~Lu,
  Nucl.\ Phys. {\bf B738}, 243 (2006);
  H.~Y.~Cheng and C.~K.~Chua,
  Phys.\ Rev.\ D {\bf 80}, 114008 (2009);
  H.~Y.~Cheng and J.~G.~Smith,
  Annu.\ Rev.\ Nucl.\ Part.\ Sci.\  {\bf 59}, 215 (2009).

\bibitem{Williamson:2006hb} 
  A.~R.~Williamson and J.~Zupan,
  Phys.\ Rev.\ D {\bf 74}, 014003 (2006);
  Phys.\ Rev. \ D {\bf 74}, 039901 (2006).

\bibitem{su3}
  C.~W.~Chiang, M.~Gronau and J.~L.~Rosner,
  Phys.\ Rev.\ D {\bf 68}, 074012 (2003);
  H.~K.~Fu, X.~G.~He and Y.~K.~Hsiao,
  Phys.\ Rev.\ D {\bf 69}, 074002 (2004);
  C.~W.~Chiang, M.~Gronau, J.~L.~Rosner and D.~A.~Suprun,
  Phys.\ Rev.\ D {\bf 70}, 034020 (2004);
  C.~W.~Chiang and Y.~F.~Zhou,
  J. High Energy Phys. 12 (2006) 027;
  H.~Y.~Cheng, C.~W.~Chiang and A.~L.~Kuo,
  Phys.\ Rev.\ D {\bf 91}, 014011 (2015).

\bibitem{Albrecht:1990am} 
  H.~Albrecht {\it et al.}  (ARGUS Collaboration),
  Phys.\ Lett.\ B {\bf 241}, 278 (1990).


\bibitem{Acciarri:1995bx} 
  M.~Acciarri {\it et al.}  (L3 Collaboration),
  Phys.\ Lett.\ B {\bf 363}, 127 (1995).

\bibitem{Richichi:1999kj} 
  S.~J.~Richichi {\it et al.}  (CLEO Collaboration),
  Phys.\ Rev.\ Lett.\  {\bf 85}, 520 (2000).

\bibitem{Chang:2004fz} 
  P.~Chang {\it et al.}  (Belle Collaboration),
  Phys.\ Rev.\ D {\bf 71}, 091106 (2005).

\bibitem{Aubert:2008fu} 
  B.~Aubert {\it et al.}  (BaBar Collaboration),
  Phys.\ Rev.\ D {\bf 78}, 011107 (2008).

\bibitem{Pal:2015ewa} 
  B.~Pal {\it et al.} (Belle Collaboration),
  Phys.\ Rev.\ D {\bf 92}, 011101 (2015).



\end{thebibliography}
\end{document}